\documentclass[twocolumn,aps,pre,reprint,showpacs,amsmath,amssymb]{revtex4-1}
\usepackage{epsfig}
\usepackage{graphicx}% Include figure files
\usepackage{dcolumn}% Align table columns on decimal point
\usepackage{bm}% bold math
\usepackage[colorlinks=true, letterpaper=ture, pdfstartview=FitV, linkcolor=red, citecolor=blue, urlcolor=blue]{hyperref}
\usepackage{multirow}
\usepackage{CJK}
\usepackage{float}

\begin{document}
\begin{CJK*}{GBK}{}
\title{Hybrid scaling mechanism of critical behavior in the overlapping critical regions of classical and quantum Yang-Lee edge singularities}

\author{Yue-Mei Sun$^{1,2}$}
\author{Wen-Jing Yu$^{1}$}
\author{Xin-Yu Wang$^{1}$}
\email{wxy@jsut.edu.cn}
\author{Liang-Jun Zhai$^{1,2}$}
\email{zhailiangjun@jsut.edu.cn}

\affiliation{$^1$The school of mathematics and physics, Jiangsu University of Technology, Changzhou 213001, China}
\affiliation{$^{2}$ The Jiangsu Key Laboratory of Clean Energy Storage and Conversion, Jiangsu University of Technology, Changzhou 213001, China}

\date{\today}

\begin{abstract}
Recently, the study of scaling behavior in Yang-Lee edge singularities (YLES) has attracted sustained attention.
However, the scaling mechanism for the overlapping critical region between classical and quantum YLES remains unclear.
In this work, we investigate this question, and a hybrid scaling mechanism is introduced to characterize the scaling behavior in the overlapping regions.
The hybrid scaling mechanism asserts that in the overlapping region the scaling behavior can be described by the scaling function for both critical regions simultaneously, and it results in a constraint on the scaling functions.
The transverse Ising chain in an imaginary longitudinal field, which exhibits $(0+1)$ dimensional (D) and $(1+1)$ D quantum YLES phase transitions at zero temperature, and $(0+0)$ D and $(1+0)$ D classical YLES phase transitions at finite temperature, is employed as a model to test this hybrid scaling mechanism.
The scaling functions in the critical regions of $(0+1)$ D and $(1+1)$ D quantum YLES  as well as $(0+0)$ D and $(1+0)$ D classical YLES of such model are systematically investigated.
Furthermore, the hybrid scaling mechanisms in overlapping critical regions, particularly between classical and quantum YLES, are thoroughly examined.
Through this study, we have established a scaling mechanism capable of describing behaviors in the overlapping critical regions between classical and quantum phase transitions, which also facilitates the extraction of quantum phase transition information from classical phase transition systems.
\end{abstract}

\maketitle

\section{Introduction}

To elucidate the fundamental nature of phase transitions, Yang and Lee pioneered an approach through the study of partition function zeros, now known as Lee-Yang zeros, in the complex plane of an external magnetic field~\cite{CNYang,Lee1952}.
It was shown that singular behaviors emerge both at the critical point with vanishing symmetry-breaking field and near the edge of Lee-Yang zeros. The latter is also referred to as the Yang-Lee edge singularity (YLES)~\cite{Fisher1978,Kortman}.
Over the past several decades, YLES has remained a subject of enduring interest in the research community, standing as one of the most representative cases where renormalization group theory has been successfully applied to critical phenomena~\cite{Kurtze1979,Fisher1980,Cardy,Uzelac1979,Kurtze1978,Wang1998,Alcantara1981,Glumac1991,Basar2021,Cardy2023}.
Although experimentally realizing the YLES has long been challenging due to its occurrence in complex parameter space,
recent advances in theoretical and experimental techniques, particularly in non-Hermitian physics~\cite{ChBinek_2001,Wei2012,Wei2015,Shen2023,Ueda2022,Wei2017,Xiao2021,Binek1998,Brandner2017,Billy2008,Linrj2023,xueprl2022,xueNC2022,xuekzs2021,Jiang2019,Bergholtz2021,Ashida2020,Yang2020,Okuma2020,Zhang2020,Borgnia2020}, have now enabled its experimental observation\cite{Gao2024,Lan2024}.
Inspired by these experimental breakthroughs, studies on YLES~\cite{Lencses2023,Johnson2023,Karsch2024,Xu2022,Dimopoulo2022,Basar2022,Basar2024,Lencses2024,Jian2021}, such as critical behavior in YLES~\cite{Gao2024,Lan2024,Lencses2023,Basar2024,Lencses2024}, and the nonequilibrium dynamics in YLES~\cite{Shen2023,Sun2025,Sunym2025entropy,Xu2025,Zhai2020}, has recently regained significant momentum.

On the other hand, the third law of thermodynamics promises a `warm' condition in nature by stating that reducing the temperature of a system to zero via a finite number of processes is impossible.
Thermal fluctuations are thus inevitably involved in any real experiment.
In the low temperature region, quantum fluctuations are also engaged.
Interplays between thermal fluctuations and quantum fluctuations have contributed plenty of intriguing phenomena
~\cite{Sachdev,Isakov2003,Witkowska,Werlang2010,Werlang2012,Vojta2003,Continentino2011,Ribeiro2024,Goldenfeld1992,Mills1993,Coleman2005,Lohneysen2007,Cui2023,Senthil2004,Gegenwart2008}.
Among them, quantum critical phenomena attract special attention~\cite{Gegenwart2008,Hertz1976,Cuccoli,Lavagna,Elstner1998,Jakubczyk2010,Jost2022,Merchant2014,Vasin2015,Sachdev2011,Coleman2005}.
Although the quantum critical point exists exactly at zero temperature, it affects a broad finite temperature region~\cite{Sachdev}.
As classical phase transitions can also happen in this finite temperature region, the classical critical region and the quantum critical region come across each other frequently~\cite{Isakov2003,Continentino2011,Merchant2014,Vasin2015}.
As a consequence, in the overlapping critical region, both the long-range entangled quantum fluctuation and the thermal fluctuation play significant roles~\cite{Merchant2014,Vasin2015}.

At low temperatures, the critical regions of quantum YLES and classical YLES overlap as well.
However, the nature of critical behavior in these overlapping regions remains an open question.
This unexplored regime presents a unique opportunity to investigate how quantum and classical criticality interact and potentially give rise to novel universal behavior at their confluence.
In recent studies, a hybrid scaling mechanism that characterizes the critical behavior within the overlapping critical region has been proposed~\cite{Yin2017,Zhai2018}. This mechanism asserts that: (i) in the overlapping critical region constructed by two critical regions, the scaling functions for both regions are applicable;
(ii) these two scaling functions have a constraint relationship containing critical information from both regions.
For instance, in the study of the non-equilibrium dynamics within the transverse Ising chain model subjected to an imaginary longitudinal field, a hybrid scaling mechanism of non-equilibrium dynamics in the overlapping critical region between $(0+1)$ dimensional (D) YLES and $(1+1)$ D ferromagnetic-paramagnetic phase transition (FPPT) has been proposed and subsequently verified numerically~\cite{Zhai2018}.
Notably, similar hybrid mechanisms have recently emerged in studies of  static scaling and non-equilibrium dynamics of localization transitions, suggesting a broader applicability of this approach to mixed criticality in diverse physical contexts~\cite{Liang2024,Sahoo2025,Bu2022,Bu2023,Sunym2024,Sunym2025}.

In this study, we employ the idea of hybrid scaling mechanism to investigate the scaling behavior in the overlapping critical region between classical and quantum YLES.
The model of a transverse Ising chain in an imaginary longitudinal field is employed. It exhibits quantum phase transitions of $(0+1)$ D YLES and $(1+1)$ D YLES at zero temperature.
At finite temperatures, it instead shows classical phase transitions of $(0+0)$ D YLES and $(1+0)$ D YLES.
By studying the static behavior of the order parameter, we verify the hybrid scaling mechanism for the overlapping critical region of $(0+1)$ D and $(1+1)$ D quantum YLES, as well as for those of $(0+0)$ D and $(1+0)$ D classical YLES.
More importantly, we also apply this hybrid scaling mechanism for the overlapping critical regions between classical YLES and quantum YLES, i.e., between $(0+0)$ D and $(0+1)$ D YLES, and between $(1+0)$ D and $(1+1)$ D YLES.
These findings are important in understanding the role of quantum fluctuations and thermal fluctuations in the overlapping critical regions constituted by the quantum and classical critical regions.
Moreover, they extend the application of hybrid scaling mechanisms to the fields of classical and quantum phase transitions, establishing a unified theoretical framework that bridges the critical phenomena across classical and quantum phase transitions.

The rest of this paper is organized as follows:
In Sec.~\ref{Sec:model}, the model of the transverse Ising chain in a imaginary longitudinal field is introduced, and the relations between the critical points of classical and quantum YLES are introduced and numerically verified.
In Sec.~\ref{Sec:HybridQauntum}, the hybrid scaling mechanism in the overlapping critical regions of quantum YLES with zero temperature is introduced and numerically verified.
In Sec.~\ref{sec:hybridclassical}, the hybrid scaling mechanism in the overlapping critical regions of classical YLES with finite temperature is introduced and numerically verified.
Then, the scaling behavior of order parameter in the critical regions of quantum YLES with finite temperature is studied,
and the hybrid scaling mechanism between the quantum and classical YLES phase transition is introduced and numerically verified in Sec.~\ref{sec:hybridquantumclassical}.
Finally, a brief summary is given in Sec.~\ref{sum}.

\section{Model and the relations between the critical points of classical and quantum YLES\label{Sec:model}}

\begin{figure*}[tbp]
  \centering
  \includegraphics[width=6 in]{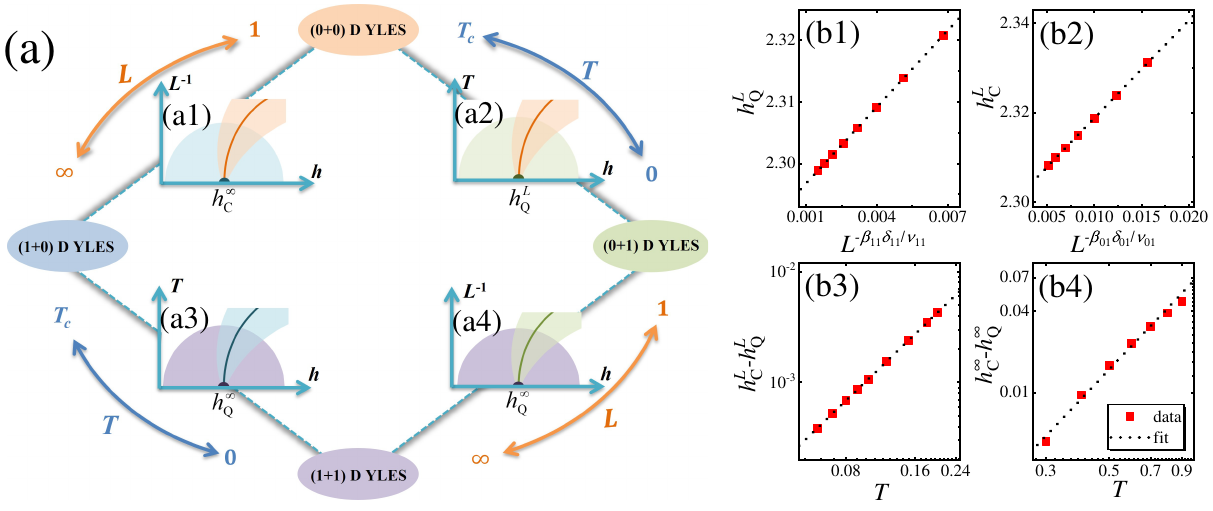}\\
  \caption{(a)The diagram of the relations between $(0+0)$ D YLES, $(1+0)$ D YLES, $(0+1)$ D YLES and $(1+1)$ D YLES.
  The light orange area represents the critical region of $(0+0)$ D YLES, the light blue area denotes the critical region of $(1+0)$ YLES, the light green area corresponds to the critical region of $(0+1)$ D YLES, and the light purple area indicates the critical region of $(1+1)$ D YLES.
  Inserts (a1), (a2), (a3), and (a4) represent the overlapping critical regions of (0+0) and (1+0) D YLES, (0+0) and (0+1) D YLES, (1+0) and (1+1) D YLES, and (0+1) and (1+1) D YLES with fixed $\lambda$, respectively.
  We take (a1) as an example to illustrate the schematic diagram of these overlapping critical regions.
  In (a1), the critical points of finite-size $(0+0)$ D YLES for different $L$ link up into a critical curve (solid orange curve), which ends at the critical point of the infinite-size $(1+0)$ D YLES. The critical region of the $(0+0)$ D YLES (light orange region) will overlap with the critical region of the $(1+0)$ D YLES (light blue area) when $L$ is sufficiently large.
   (b1)$h_{\rm Q}^L$ versus $L^{-\beta_{11}\delta_{11}/\nu_{11}}$ for $\lambda=5$ and $T=0$.
  (b2)$h_{\rm C}^L$ versus $L^{-\beta_{01}\delta_{01}/\nu_{01}}$ for $\lambda=5$ and $T=0.3$.
  (b3)Log-log plot of $h_{\rm C}^L-h_{\rm Q}^L$ versus $T$ for $\lambda=5$ and $L=10$.
  (b4)Log-log plot of $h_{\rm C}^{\infty}-h_{\rm Q}^{\infty}$ as a function of $T$ for $\lambda=5$.}
  \label{fig:FitSketch}
\end{figure*}

%For this model, the ordinary ferromagnetic-paramagnetic phase transition (FPPT) occurs at the critical point of $(\lambda,h)=(1,0)$.
%Besides this FPPT critical point,
\begin{table*}
  \centering
  \caption{Critical exponents for the $(0+0)$ D, $(0+1)$ D, $(1+0)$ D and the $(1+1)$ D universality class, respectively~\cite{Uzelac1979,Fisher1978,Yin2017,Zhai2018}.}
    \begin{tabular}{c c c c c}
%    \hline
    \hline
      \multicolumn{1}{c|}{universality class } &$\nu$   &$\beta$ & $\delta$  & $z$ \\
      \hline
    \multicolumn{1}{c|}{$(0+0)$ D } &$\nu_{00}=-0.62$   &$\beta_{00}=1$ & $\delta_{00}=-1$  & $z_{00}$=1.633 \\
    \multicolumn{1}{c|}{$(0+1)/(1+0)$ D } &$\nu_{01} =-1$   &$\beta_{01}=1$ & $\delta_{01}=-2$  & $z_{01}=1$ \\
   % \multicolumn{1}{c|}{$(1+0)$ D } &$\nu_{01} =-1$   &$\beta_{01}=1$ & $\delta_{01}=-2$  & $z_{01}=1$ \\
    \multicolumn{1}{c|}{$(1+1)$ D } &$\nu_{11}=-5/2$  & $\beta_{11}=1$ & $\delta_{11}=-6$  & $z_{11}=1$ \\
    %\multicolumn{1}{c|}{Tricritical Heisenberg} &0.5   &0.25 & 1  & 5  & 0 & 0 \\
    \hline
    \end{tabular}%
  \label{tabexp}%
\end{table*}%

\subsection {model and the order parameter}
To illustrate our scaling mechanism, we take a model of the transverse Ising chain in an imaginary longitudinal field as an example~\cite{Uzelac1979,Uzelac1980}.
The Hamiltonian reads
\begin{equation}
\mathcal{H}=-\sum_{n=1}^{L}{\sigma_n^z\sigma_{n+1}^z}-\lambda\sum_{n=1}^L\sigma_n^x-i h \sum_{n=1}^L\sigma_n^z,
\label{model}
\end{equation}
in which $\sigma_n^{x,z}$ are the Pauli matrices in $x$ and $z$ directions at the $n$ site respectively, $\lambda$ is the transverse field, $h$ is the imaginary longitudinal field, and $L$ is the lattice size.
Note that Eq.~(\ref{model}) is a non-Hermitian model.
The reason for choosing this model is that the critical phenomena in model~(\ref{model}) are very similar to the usual Hermitian ones, e.g., the ordinary FPPT of this model occurs at the critical point of $(\lambda,h)=(1,0)$~\cite{Sachdev},
but the YLES therein can occur at finite-size systems~\cite{Uzelac1980,Yin2017,Zhai2018}.

At zero temperature, this model with small and medium sizes has critical points for the YLES at $(\lambda,h)=(\lambda_{\rm Q}^L,h_{\rm Q}^L)$ with $\lambda>1$~\cite{Fisher1980}, where the superscript capital $L$ denotes the phase transition occurring at finite lattice size $L$.
The YLES near $(\lambda_{\rm Q}^L,h_{\rm Q}^L)$ in model (\ref{model}) belongs to the $(0+1)$ D universality class.
Although the Hamiltonian $\mathcal H$ is non-Hermitian, the emergence of YLES signifies the vanishing of the energy gap~\cite{Gehlen1991}, and unlike conventional phase transitions confined to the thermodynamic limit~\cite{Sachdev}, the $(\lambda_{\rm Q}^L,h_{\rm Q}^L)$ can manifest in systems of finite sizes as well~\cite{Kortman}.
In the thermodynamic limit of $L=\infty$, this model shows a $(1+1)$ D YLES phase transition at $(\lambda,h)=(\lambda_{\rm Q}^{\infty},h_{\rm Q}^{\infty})$, and the YLES near $(\lambda_{\rm Q}^{\infty},h_{\rm Q}^{\infty})$  belongs to the $(1+1)$ D universality class~\cite{Fisher1978}.

For finite temperature, the model~(\ref{model}) of small and medium sizes will exhibit classical YLES at the phase transition point $(\lambda,h)=(\lambda_{\rm C}^L,h_{\rm C}^L)$, similar to the quantum YLES phase transition, and the YLES near $(\lambda_{\rm C}^L,h_{\rm C}^L)$ belongs to the $(0+0)$ D universality class.
Similarly, as the size of the system gradually increases to $L=\infty$, a $(1+0)$ D YLES will also occur at $(\lambda,h)=(\lambda_{\rm C}^{\infty},h_{\rm C}^{\infty})$, and the YLES near $(\lambda_{\rm C}^{\infty},h_{\rm C}^{\infty})$ belongs to the $(1+0)$ D universality class~\cite{Fisher1978,Kurtze1979}.

As illustrated in Fig.~\ref{fig:FitSketch}, we demonstrate the relationship between quantum and classical YLES in model Eq.~(\ref{model}).
With increasing system size, the critical point of $(0+1)$ D YLES approaches that of $(1+1)$ D YLES, while the critical point of $(0+0)$ D YLES similarly converges toward that of (1+0)D YLES.
Additionally, as $L$ increases, the critical regions of both $(0+0)$ D YLES and $(0+1)$ D YLES also decrease.
Consequently, for sufficiently large $L$, two distinct overlapping critical regions emerge: (i) between $(0+0)$ D and $(1+0)$ D YLES at the same temperature $T$, and (ii) between $(0+1)$ D and $(1+1)$ D YLES at zero temperature.
As shown in the inserts (a1) and (a4) of Fig.~\ref{fig:FitSketch}, we present schematic diagrams of the overlapping critical regions between $(0+0)$ D and $(1+0)$ D YLES, as well as between $(0+1)$ D and $(1+1)$ D YLES.

On the other hand, as the temperature gradually decreases, the phase transition point of $(0+0)$ D YLES and that of $(0+1)$ D YLES will gradually approach each other, while the phase transition point of $(1+0)$ D YLES and that of $(1+1)$ D YLES will also gradually converge.
Consequently, in the low temperature region, the critical regions of $(0+0)$ D YLES and $(0+1)$ D YLES will overlap, and similarly, the critical regions of $(1+0)$ D YLES and $(1+1)$ D YLES will also exhibit overlapping behavior.
As shown in the inserts (a2) and (a3) of Fig.~\ref{fig:FitSketch}, we present schematic diagrams of the overlapping critical regions between $(0+0)$ D and $(0+1)$ D YLES, as well as between $(1+0)$ D and $(1+1)$ D YLES.

The order parameter $M$ is defined as~\cite{Pathria2024}
\begin{equation}\label{parameter}
 M={\rm Im}[{\rm Tr}(\rho_E\hat{M})],
\end{equation}
in which $\hat{M}=\sum_n^L{\sigma_{n}^z}/L$.
At $T=0$, $M$ becomes~\cite{Yin2017,Gehlen1991}
\begin{equation}\label{parameterQ}
 M=\frac {{\rm Im}{\langle\Psi_g^L|\hat{M}|\Psi_g^R\rangle}} {{\langle\Psi_g^L|\Psi_g^R\rangle}},
\end{equation}
in which $\langle \psi_g^L|$ and $\psi_g^R\rangle$ are
the normalized left and right eigenvectors satisfying
$\mathcal{H}|\psi^R_g\rangle=E_g|\psi^R_g\rangle$ and $\langle\psi^L_g| \mathcal{H}=\langle\psi^L_g| E_g$, and $E_g$ is eigenenergy with lowest real part.
At the YLES phase transition point, $M$ diverges.

In the following discussion, we will take the scaling behavior of $M$ with respect to $h$ under constant $\lambda$ as an example to investigate the hybrid scaling mechanism in the overlapping critical regions.
We choose the value of $\lambda$ far away from the critical point of the $(1+1)$ D FPPT, and only discuss the overlap between the critical regions of classical and quantum YELS phase transitions.

\subsection{relations between the classical and quantum YLES critical points}
Fig.~\ref{fig:FitSketch}(a) demonstrates that the critical points of quantum and classical YLES can be connected through both temperature and size dimensions.
We now proceed to discuss the relations between classical and quantum YLES critical points under temperature and size variations, respectively.

We first investigate the relations between the critical points when varying the system size, that is, the relation between quantum YLES phase transition points at zero temperature, along with the relation between classical YLES critical points at the same finite temperatures.
At $T=0$, the phase transition points of the $(0+1)$ D and $(1+1)$ D YLES, $h_{\rm Q}^{L}$ and $h_{\rm Q}^{\infty}$ for a given $\lambda$ satisfy~\cite{Gehlen1991,Yin2017}
\begin{eqnarray}
h_{\rm Q}^{L}-  h_{\rm Q}^{\infty} &=& C_1(\lambda)L^{-\frac {\beta_{11}\delta_{11}} {\nu_{11}}},
\label{Relation01_11}
\end{eqnarray}
where $\beta_{11}$, $\delta_{11}$, and $\nu_{11}$ are the critical exponents for the $(1+1)$ D universality class, and $C_i(\lambda)$ is a dimensionless function.
For the sake of clarity, we list all the relevant exponents in Table \ref{tabexp}.
Similarly, at the same temperatures, the phase transition points of the $(0+0)$ D and $(1+0)$ D YLES for a given $\lambda$, $h_{\rm C}^{L}$ and $h_{\rm C}^{\infty}$ satisfy
\begin{eqnarray}
  h_{\rm C}^{L}-h_{\rm C}^{\infty} &=& C_2(\lambda) L^{\frac {-\beta_{01}\delta_{01}} {\nu_{01}}},
  \label{Relation00_10}
\end{eqnarray}
where $\beta_{01}$, $\delta_{01}$, and $\nu_{01}$ are the critical exponents for the $(1+0)$ D universality class.

We next investigate the relationships between the critical points as the temperature varies, that is, the relationships between the critical points of classical and quantum YLES.
In principle, the critical behavior can be extracted from the density matrix $\rho_E\equiv {\rm exp}(-\mathcal H/T)/Z$, in which $Z$ is the partition function.
From this formula, one finds that the roles played by the temperature are twofold: in the quantum criticality, the temperature is an inverse of the imaginary-time length with an anisotropic exponent $z_{\rm q}$; while in the classical criticality, the reduced relative temperature, $\tau\equiv (T-T_c)/T_c$ with $T_c$ being the classical critical point, drives the classical phase transition~\cite{Sachdev,Hertz1976,Vojta2003,Goldenfeld1992,Sachdev2011,Mills1993}.
Near the zero temperature, this duplex effect dictates that the critical behavior in the low temperature classical critical region is also affected by the quantum critical point.
This leads to a scaling relation between classical critical points and temperature ~\cite{Coleman2005,Lohneysen2007,Cui2023,Senthil2004,Gegenwart2008}.

For model (\ref{model}), the critical points of $(0+0)$ D YLES and $(0+1)$ D YLES with the same $L$, satisfy
\begin{eqnarray}
  h_{\rm C}^{L}-h_{\rm Q}^{L} &=& C_3(\lambda) T^{\frac {\beta_{01}\delta_{01}} {\nu_{01}z_{01}}}.
  \label{Relation00_01}
\end{eqnarray}
where $z_{01}$ is a critical exponent for the $(0+1)$ D universality class.
While for the $(1+0)$ D YLES and $(1+1)$ D YLES, the critical points are related by
\begin{eqnarray}
  h_{\rm C}^{\infty}-h_{\rm Q}^{\infty} &=& C_4(\lambda) T^{\frac {\beta_{11}\delta_{11}} {\nu_{11}z_{11}}},
  \label{Relation10_11}
\end{eqnarray}
where $z_{11}$ is a critical exponent for the $(1+1)$ D universality class.

Here, we locate the critical positions of quantum and classical YLES phase transitions by numerically calculating the divergence points of $M$, thereby verifying the relationships mentioned above.
In Fig.~\ref{fig:FitSketch} (b1),
Eq.~(\ref{Relation01_11}) is numerically verified by plotting $h_{\rm Q}^L$ versus $L^{-\beta_{11}\delta_{11}/\nu_{11}}$ for $\lambda=5$.
$h_{\rm Q}^L$ versus $L^{-\beta_{11}\delta_{11}/\nu_{11}}$ is a straight line demonstrating that $h_{\rm Q}^{L}\propto L^{-\beta_{11}\delta_{11}/\nu_{11}}$, confirming Eq.~(\ref{Relation01_11}).
Then, by studying the relationship between $ h_{\rm C}^{L}$ and $L^{-\beta_{01}\delta_{01}/\nu_{01}}$ for $\lambda=5$, we have verified Eq.~(\ref{Relation00_10}).
As shown in Fig.~\ref{fig:FitSketch} (b2),
the plot of $ h_{\rm C}^{L}$ versus $L^{-\beta_{01}\delta_{01}/\nu_{01}}$ is a straight line, confirming the linear relationship between the two and also validating Eq.~(\ref{Relation00_10}).

To verify Eq.~(\ref{Relation00_01}), we first study the analytical expression of $M$ with $L=1$.
For a single-spin system, there is no ferromagnetic coupling in model~(\ref{model}), and $M$ at finite temperature can be exactly solved as
\begin{eqnarray}
  M(T) &=& -{\rm Im}[\frac{ih}{\sqrt{\lambda^2-h^2}}\tanh(\frac {\sqrt{\lambda^2-h^2}} {T})].
\end{eqnarray}
We find that the divergence of $M$ can be classified into two classes:
(i) $M$ diverges at $h=\pm\lambda$, which is independent of temperature, and it corresponds to the critical points of the $(0+1)$ D YLES at $T=0$.
(ii) $M$ also diverges at $h=h_{\rm C}^{L}=\pm\sqrt{\lambda^2+\pi^2T^2/4}$, which corresponds to the critical points of the $(0+0)$ D YLES.
One finds that $h_{\rm C}^{L}$ and $h_{\rm Q}^{L}$ become identical as $T\rightarrow0$, and the two satisfy $h_{\rm C}^{L}-h_{\rm Q}^{L} \propto T^{2}$ confirming Eq.~(\ref{Relation00_01}).
For larger systems, we numerically determine the values of $h_{\rm C}^{L}$ at different temperatures and $h_{\rm Q}^{L}$ at zero temperature with a given $\lambda$ and $L$.
Then we fit the curve of $h_{\rm C}^{L}-h_{\rm Q}^{L}$ versus $T$, as shown in Fig.~\ref{fig:FitSketch}(b3), obtaining an exponent of $1.9971$, which is close to ${\beta_{01}\delta_{01}}/{\nu_{01}z_{01}}=2$, thereby validating Eq.~(\ref{Relation00_01}).

To validate Eq.~(\ref{Relation10_11}), we use numerically obtained values of $h_{\rm Q}^{L}$ for different $L$ to fit $h_{\rm Q}^{\infty}$ at $T=0$,
and use the numerically obtained values of $h_{\rm C}^{L}$ for different $L$ to fit $h_{\rm C}^{\infty}$ at different temperatures.
Subsequently, we fit the exponential relationship between $h_{\rm C}^{\infty}-h_{\rm Q}^{\infty}$ and $T$, as shown in Fig.~\ref{fig:FitSketch}(b4).
Power law fitting yields an exponent of $2.2311$, which is close to $\beta_{11}\delta_{11}/\nu_{11}z_{11}=2.4$, confirming Eq.~(\ref{Relation10_11}).

\section{Hybridized scaling mechanism for the quantum YLES at $T=0$\label{Sec:HybridQauntum}}
In this section, we study the scaling mechanism of $M$ with respect to the variation of $h$ in the critical regions of $(0+1)$ D and $(1+1)$ D YLES at $T=0$, and the hybridized scaling mechanism in the overlapping regions.
According to the hybrid scaling mechanism, firstly, in this overlapping critical region, the scaling functions followed by $(0+1)$ D and $(1+1)$ D YLES should remain valid;
Secondly, there is a the hybrid scaling function related to the critical index constraint relationship between the two, which contains both $(0+1)$ D and $(1+1)$ D YLES critical information.

\begin{figure}
  \centering
  \includegraphics[width=2.5 in]{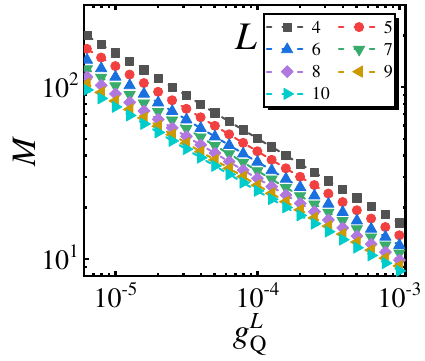}\\
  \caption{$M$ versus $g_{\rm Q}^L$ at $T=0$ for different $L$ are plotted.
  Here, we use $\lambda=5$, and $h_{\rm Q}^L=2.433731$, $2.377299$, $2.348062$, $2.331236$, $2.320787$, $2.313911$, $2.309176$ for
  $L=4, 5, 6, 7, 8,  9, 10$, respectively.
  The double-logarithmic coordinate is used.
  }\label{scale01T0}
\end{figure}

In the region around the critical point $(\lambda_{\rm Q}^{L},h_{\rm Q}^{L})$ of the $(0+1)$ D YLES, $M$ with fixed $\lambda$ and $T=0$ exhibits a divergence that scales as~\cite{Fisher1978,Zhai2018}
\begin{eqnarray}
  M \sim (g_{\rm Q}^{L})^{\frac{1}{\delta_{01}}},
  \label{Eq:scaling01T0}
\end{eqnarray}
where $g_{\rm Q}^{L}=h-h_{\rm Q}^{L}$ is the distance to the critical point.
As shown in Fig.~\ref{scale01T0}, we present the plots of $M$ versus $g_{\rm Q}^L$ for different system sizes $L$.
It can be seen that the curves for different $L$ are parallel straight lines in a double-logarithmic coordinate system.
Moreover, power-law fitting reveals that the average slope of these curves is $-0.4983$ close to the theoretical value of $-1/2$,  thereby confirming the validity of Eq.~(\ref{Eq:scaling01T0}).
Additionally, these results indicate that, regardless of whether the system size is small or moderate, the behavior of $M$ near the $(\lambda_{\rm Q}^{L},h_{\rm Q}^{L})$ can be described by the $(0+1)$ D YLES scaling function.

\begin{figure}
  \centering
  \includegraphics[width=3.5 in]{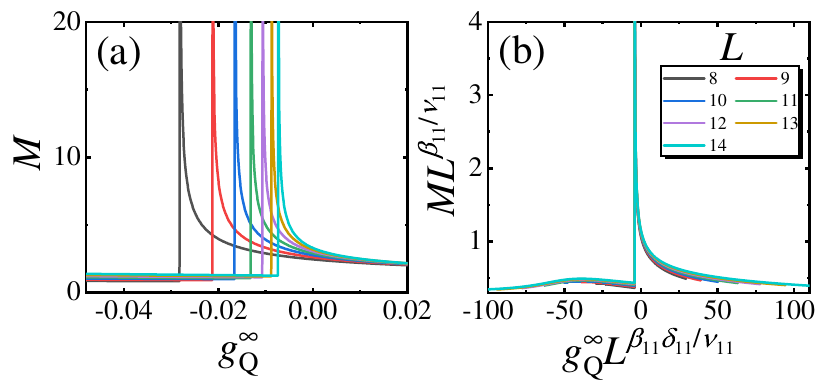}\\
  \caption{(a)$M$ versus $g_{\rm Q}^\infty$ for different $L$ are plotted.
  (b) The rescaled curves according to Eq.~(\ref{Eq:scaling11T0}).
  Here, we use $\lambda=5$, and $h_{\rm Q}^{\infty}=2.292657$.}\label{scale11T0}
\end{figure}

\begin{figure}
  \centering
  \includegraphics[width=3.5 in]{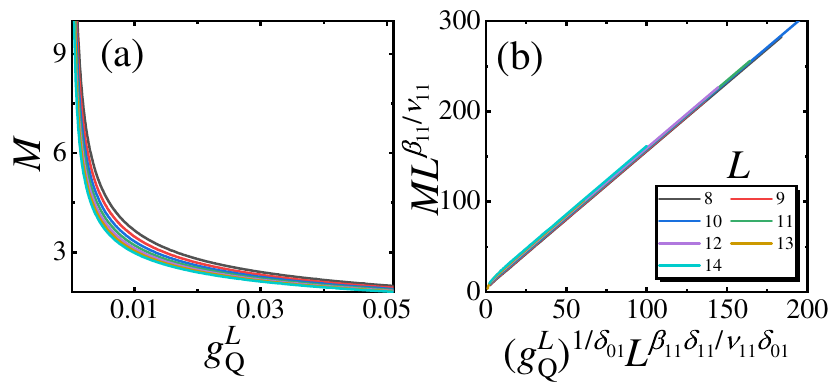}\\
  \caption{(a) $M$ versus $g_{\rm Q}^L$ for different $L$ at $T=0$.
  (b) The rescaled cures according to Eq.~(\ref{constraint01-11T0}).
  Here, we use $\lambda=5$, and the critical points of $(0+1)$ D YLES are $h_{\rm Q}^L=2.320787, 2.313911, 2.309176, 2.305794, 2.303305, 2.301426$ and $2.299978$ for $L=8, 9, 10, 11, 12, 13$ and $14$, respectively.}\label{Fig:constrains0111}
\end{figure}

For medium-sized systems, the critical regions for the $(0+1)$ D and $(1+1)$ D YLESs overlap with each other.
In this overlap region, besides Eq.~(\ref{Eq:scaling01T0}), the behavior of $M$ should also satisfy the $(1+1)$ D YLES scaling functions ~\cite{Yin2017}.
By taking into account of the finite-size correction, $M$ near the critical point of $(1+1)$ D YLES at $T=0$ satisfies,
\begin{eqnarray}
  M &=& L^{-\frac{\beta_{11}}{\nu_{11}}}f_1[g_{\rm Q}^{\infty}L^{\frac{\beta_{11}\delta_{11}}{\nu_{11}}}],
  \label{Eq:scaling11T0}
\end{eqnarray}
where $g_{\rm Q}^{\infty}=h-h_{\rm Q}^{\infty}$, and $f_i$ is the scaling function.

As depicted in Fig.~\ref{scale11T0}, we have calculated the variation of $M$ with respect to $g_{\rm Q}^{\infty}$ in the critical region of the $(1+1)$ D YLES for medium-sized systems at $T=0$, as well as the rescaled curves obtained from Eq.~(\ref{Eq:scaling11T0}).
Evidently, the rescaled curves obtained from Eq.~(\ref{Eq:scaling11T0}) collapse onto a single line, as shown in Fig.~\ref{scale11T0} (b), thereby validating the correctness of Eq.~(\ref{Eq:scaling11T0}).
As can be observed from the results presented in Fig.~\ref{scale01T0} and Fig.~\ref{scale11T0}, the scaling functions for $(0+1)$ D and $(1+1)$ D YLES are simultaneously applicable in medium-sized systems.
This indicates the existence of an overlapping critical region in medium-sized systems. In this region, the scaling functions for the two phase transitions of $(0+1)$ D and $(1+1)$ D YLES apply simultaneously, thus validating the first assumption of the hybrid scaling mechanism.

Within this overlapping critical region, according to the hybridized scaling mechanism, there must exist a constraint between the scaling equations of Eq. (\ref{Eq:scaling01T0}) and (\ref{Eq:scaling11T0}).
To obtain the constraint equations, we substitute Eq.~(\ref{Eq:scaling11T0}) into Eq.~(\ref{Eq:scaling01T0}), and the following equation is derived
\begin{eqnarray}
  M &=&L^{\frac{\beta_{11}}{\nu_{11}}(\frac{\delta_{11}}{\delta_{01}}-1)}(g_{\rm Q}^L)^{\frac{1}{\delta_{01}}}.
  \label{constraint01-11T0}
\end{eqnarray}
%L^{-\frac{\beta_{11}}{\nu_{11}}}[g_{\rm YL}^LL^{\frac{\beta_{11}\delta_{11}}{\nu_{11}}}]^{1/\delta_{01}}
One finds that Eq.~(\ref{constraint01-11T0}) contains the critical exponents for both the $(0+1)$ D and $(1+1)$ D YLES, which can reflect the relationship between Eq.~(\ref{Eq:scaling11T0}) and Eq.~(\ref{Eq:scaling01T0}).
Therefore, Eq.~(\ref{constraint01-11T0}) can be regarded as the constraint between $(0+1)$ D and $(1+1)$ D YLES scaling functions.

In Fig.~\ref{Fig:constrains0111}, the constraint between the $(0+1)$ D YLES and $(1+1)$ D YLES is verified by validating Eq.~(\ref{constraint01-11T0}).
In Fig.~\ref{Fig:constrains0111} (a), the curves of $g_{\rm Q}^L$ dependence of $M$ of different $L$ at $T=0$ are plotted.
After rescaling according to Eq.~(\ref{constraint01-11T0}), these curves all collapse into one curve, thereby proving Eq.~(\ref{constraint01-11T0}).
This proves the second assumption of the hybrid scaling mechanism.

Combining the results from Figs.~\ref{scale01T0} to~\ref{Fig:constrains0111}, we have  verified the applicability of the hybrid scaling mechanism within the overlapping critical regions of quantum YLES.
In fact, previous study has also presented a hybrid Kibble-Zurek mechanism (HKZM) for non-equilibrium dynamics within the overlapping critical regions of $(0+1)$ D and $(1+1)$ D YLES~\cite{Yin2017}, which can be considered a form of hybrid scaling mechanism.
Our results, on the other hand, pertain to the hybrid scaling mechanism for static behavior.

\section{Hybridized scaling mechansim for the classical YLES at finite temperature\label{sec:hybridclassical}}
In this section, we study the scaling behavior in the critical regions of $(0+0)$ D and $(1+0)$ D YLES at finite temperature, as well as the hybridized scaling mechanism in the overlapping critical regions.

At finite temperature, $M$ will no longer diverge at the critical points of quantum YLES, but will diverge at the critical points of classical YLES.
Therefore, in the critical region of $(0+0)$ D YLES, when $\lambda$ remains constant, the variation of $M$ follows the following scaling function~\cite{Fisher1978,Gehlen1991}
\begin{eqnarray}
  M \sim (g_{\rm C}^{L})^{\frac{1}{\delta_{00}}},
  \label{Eq:scaling00}
\end{eqnarray}
in which $g_{\rm C}^{L}=h-h_{\rm C}^{L}$.
It should be noted that although Eq.~(\ref{Eq:scaling00}) has the same form as Eq.~(\ref{Eq:scaling01T0}), the values of $\delta_{00}$ and $\delta_{01}$ are different.
This means that $M$ will diverge with different exponents.

\begin{figure}
  \centering
  \includegraphics[width=2.5 in]{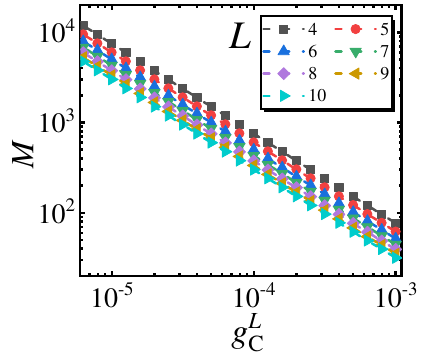}
  \caption{$M$ versus $g_{\rm C}^{L}$ in the log-log plot.
  Here, we use $\lambda=5$ and $T=0.3$. The critical points of the $(0+0)$ D YLES are $h_{\rm C}^{L}=2.447562$, $2.389941$, $2.359806$, $2.342268$, $2.331231$, $2.323860$, and $2.318697$ for $L=4$, $5$, $6$, $7$, $8$, $9$, and $10$, respectively.
  The double-logarithmic coordinate is used.}
  \label{sacle00}
\end{figure}

In Fig.~\ref{sacle00}, the dependence of $M$ on $g_{\rm C}^L$ is plotted for different values of $L$, with $\lambda=5$ and $T=0.3$.
It is observed that near each $h_{\rm C}^{L}$, the curves of $M$ versus $g_{\rm C}^{L}$ appear as parallel straight lines in double logarithmic coordinates.
The averaged slope obtained from power-law fitting is $-0.9991$, which is remarkably close to the theoretical prediction of $-1$.
These results validate Eq.~(\ref{Eq:scaling00}), and also indicate that the scaling functions of $(0+0)$ D YLES are valid for small or medium-sized systems.

For medium-sized systems, in the overlapping region constituted by the critical regions of $(0+0)$ D and $(1+0)$ D YLES, the scaling function of $(1+0)$ D YLES is also applicable.
Near the critical points of $(1+0)$ D YLES, after considering the corrections due to finite-size effects, the variation of $M$ for fixed $\lambda$ reads,
\begin{eqnarray}
  M &=& L^{-\frac{\beta_{01}}{\nu_{01}}}f_2[g_{\rm C}^{\infty}L^{\frac{\beta_{01}\delta_{01}}{\nu_{01}}}],
  \label{Eq:scaling10}
\end{eqnarray}
where $g_{\rm C}^{\infty}=h-h_{\rm C}^{\infty}$.

\begin{figure}
  \centering
  \includegraphics[width=3.5 in]{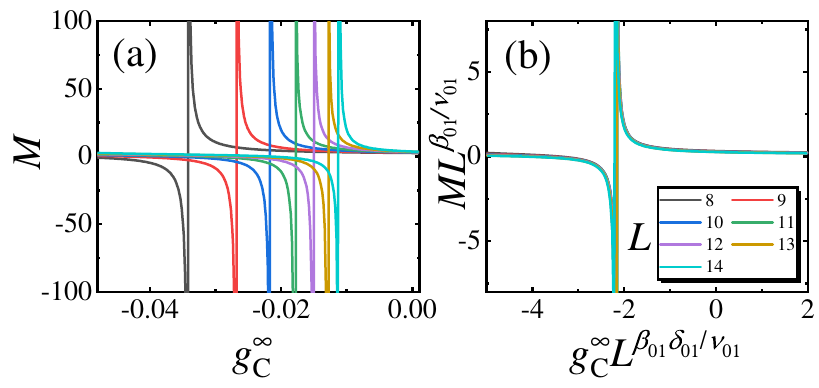}\\
  \caption{(a) $M$ as a function of $g_{\rm C}^{\infty}$ for different $L$.
  (b) The rescaled curves of $ML^{\beta_{01}/\nu_{01}}$ versus $g_{\rm C}^{\infty}L^{\beta_{01}\delta_{01}/\nu_{01}}$.
  Here, we use $\lambda=5$, $T=0.3$, and $h_{\rm C}^{\infty}$ is $2.297034$.}\label{sacle10}
\end{figure}

To verify the scaling function of Eq.~(\ref{Eq:scaling10}),
the $g_{\rm C}^{\infty}$ dependence of $M$ for different $L$ are plotted in Fig.~\ref{sacle10} (a).
After rescaling $M$ and $g_{\rm C}^{\infty}$ according to Eq.~(\ref{Eq:scaling10}), these rescaled curves collapse onto each other as plotted in Fig.~\ref{sacle10} (b), confirming Eq.~(\ref{Eq:scaling10}).
\begin{figure}
  \centering
  \includegraphics[width=3.5 in]{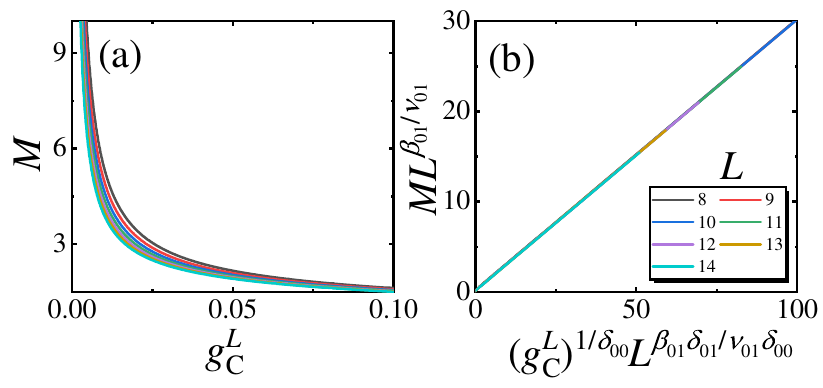}\\
  \caption{(a) $M$ versus $g_{\rm C}^{L}$ for different $L$ at $T=0.3$.
  (b) The rescaled cures according to Eq.~(\ref{constraint00-10}).
  Here, we use $\lambda=5$, and the critical points of $(0+0)$ D YLES are $h_{\rm C}^{L}=2.331231, 2.323860, 2.318697, 2.314940, 2.312118, 2.309939$ and $2.308218$ for $L=8, 9, 10, 11, 12, 13$ and $14$, respectively.}
  \label{Fig:constrains0010}
\end{figure}

Similarly, in the overlapping critical region constructed by the critical regions of $(0+0)$ D YLES and $(1+0)$ D YLES, the constraints between the scaling functions can be obtained by substituting Eq.~(\ref{Eq:scaling10}) into Eq.~(\ref{Eq:scaling00}), which reads
\begin{eqnarray}
  M &=&L^{\frac{\beta_{01}}{\nu_{01}}(\frac{\delta_{01}}{\delta_{00}}-1)}(g_{\rm C}^{L})^{\frac{1}{\delta_{00}}}.
  \label{constraint00-10}
\end{eqnarray}
To verify the constraint between the $(0+0)$ D YLES and $(1+0)$ D YLES, plots of $M$ versus $g_{\rm CL}^{L}$ for different $L$ are shown at $T=0.3$
in Fig.~\ref{Fig:constrains0010} (a).
After rescaling according to Eq.~(\ref{constraint00-10}), these curves all collapse into one curve, confirming Eq.~(\ref{constraint00-10}).

Indeed, the calculations presented in Figs.~\ref{sacle00} to~\ref{Fig:constrains0010} also confirm the applicability of the hybrid scaling mechanism within the overlapping critical regions of classical YLES at finite temperatures.
These conclusions further extend the application of the hybrid scaling mechanism to classical phase transitions.

\section{Hybridized scaling functions between the classical YLES and quantum YLES\label{sec:hybridquantumclassical}}
In this section, we will discuss the hybridized scaling mechanisms between classical YLES and quantum YLES, that is, the scaling functions in the overlapping critical regions of $(0+0)$ D and $(0+1)$ D YLES, as well as the critical regions of $(1+0)$ and $(1+1)$ D YLES.
\subsection{the $(0+1)$ D YLES and $(1+1)$ D YLES scaling functions at low temperatures}
We investigated the critical behavior of $M$ near the quantum critical point at low temperatures.
At finite temperatures, while $M$ does not diverge at quantum YLES critical points, these points significantly influence the scaling behavior of $M$.
Therefore, by taking $T$ as a relevant variable, in the vicinity of the critical point $(\lambda_{\rm Q}^{L},h_{\rm Q}^{L})$, the scaling behavior of $M$ with a fixed $\lambda$ is characterized by
\begin{eqnarray}
  M &=& T^{\frac{\beta_{01}}{\nu_{01}z_{01}}}f_{3}[g_{\rm Q}^L T^{-\frac{\beta_{01}\delta_{01}}{\nu_{01}z_{01}}}].
  \label{Eq:scaling01}
\end{eqnarray}
Note that there is no finite-size effect correlation in Eq.~(\ref{Eq:scaling01}), since $L$ is irrelevant in this $(0+1)$ D YLES.

Similarly, at low temperatures, $M$ near $(\lambda_{\rm Q}^{\infty},h_{\rm Q}^{\infty})$ satisfies the following scaling function
\begin{eqnarray}
\label{Eq:scaling11}
  M= L^{-\frac{\beta_{11}}{\nu_{11}}}f_{4}[g_{\rm Q}^{\infty}T^{-\frac{\beta_{11}\delta_{11}}{\nu_{11}z_{11}}}, g_{\rm Q}^{\infty}L^{\frac {\beta_{11}\delta_{11}}{\nu_{11}}}].
\end{eqnarray}

\begin{figure}
  \centering
  \includegraphics[width=3.45 in]{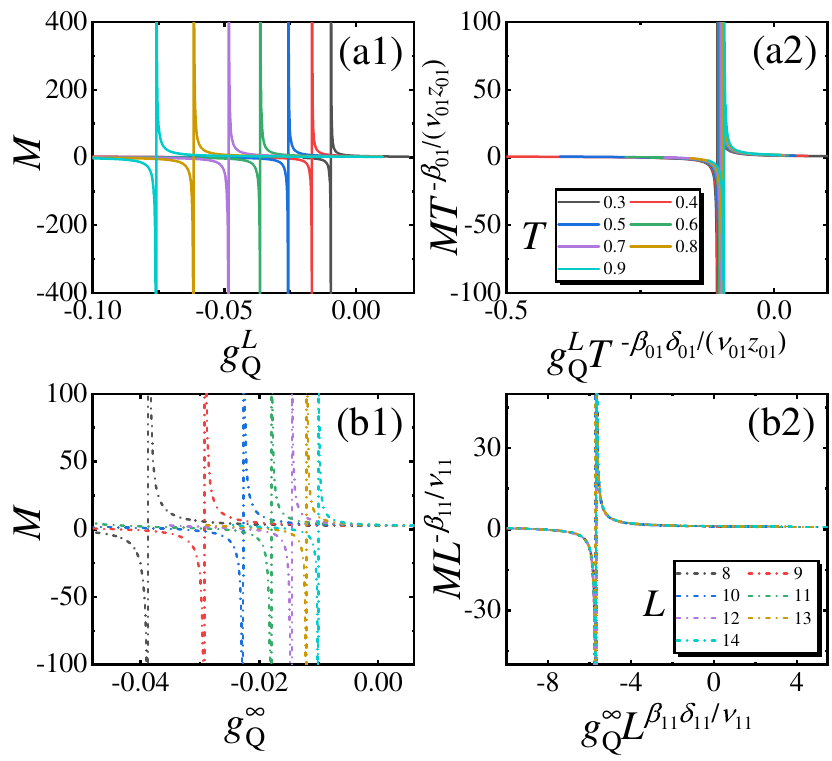}\\
  \caption{(a1) $M$ versus $g_{\rm Q}^L$ for different $T$ are plotted with $L=10$.
   (a2) The rescaled curves according to Eq.~(\ref{Eq:scaling01}) with $L=10$.
   (b1)$M$ versus $g_{\rm Q}^\infty$ for different $L$ are plotted.
  (b2) The rescaled curves according to Eq.~(\ref{Eq:scaling11}).
   Here, we use $\lambda=5$, and $h_{\rm Q}^L=2.309176$ for $L=10$ in (a1) and (a2),
  and $\lambda=5$, $h_{\rm Q}^{\infty}=2.292657$, $LT^{1/z_{11}}=2.4$ in (b1) and (b2).}\label{scale0111T}
\end{figure}

In Figs.~\ref{scale0111T} (a1) and (a2), the scaling function of $(0+1)$ D YLES is tested for medium-sized lattices for different temperatures.
It can be observed that the curves of $M$ versus $g_{\rm Q}^L$ for medium sizes collapse very well after being rescaled according to Eq.~(\ref{Eq:scaling01}).

To verify the scaling function of Eq.~(\ref{Eq:scaling11}), we rewrite $g_{\rm Q}^{\infty}T^{-{\beta_{11}\delta_{11}}/{\nu_{11}z_{11}}}$ as $g_{\rm Q}^{\infty}L^{ {\beta_{11}\delta_{11}}/\nu_{11}}(LT^{1/z_{11}})^{-{\beta_{11}\delta_{11}}/\nu_{11}}$.
Therefore, when we fix $LT^{1/z_{11}}$ as a constant, the two variables in Eq.~(\ref{Eq:scaling11}) can be regarded as a single one.
In Figs.~\ref{scale0111T} (b1) and (b2), the scaling function of $(1+1)$ D YLES is verified by fixing $LT^{1/z_{11}}$ to a random constant for different $L$.
It is shown that after rescaling according to Eq.~(\ref{Eq:scaling11}), the rescaled curves match with each other very well confirming Eq.~(\ref{Eq:scaling11}).

Therefore, based on the numerical results from Fig.~\ref{scale0111T}, we have demonstrated that the scaling functions of $(0+1)$ D and $(1+1)$ D quantum YLES remain valid at low temperatures.

\subsection{Hybridized scaling mechanism in the overlapping regions between classical and quantum YLES}
\begin{figure}
  \centering
  \includegraphics[width=3.6 in]{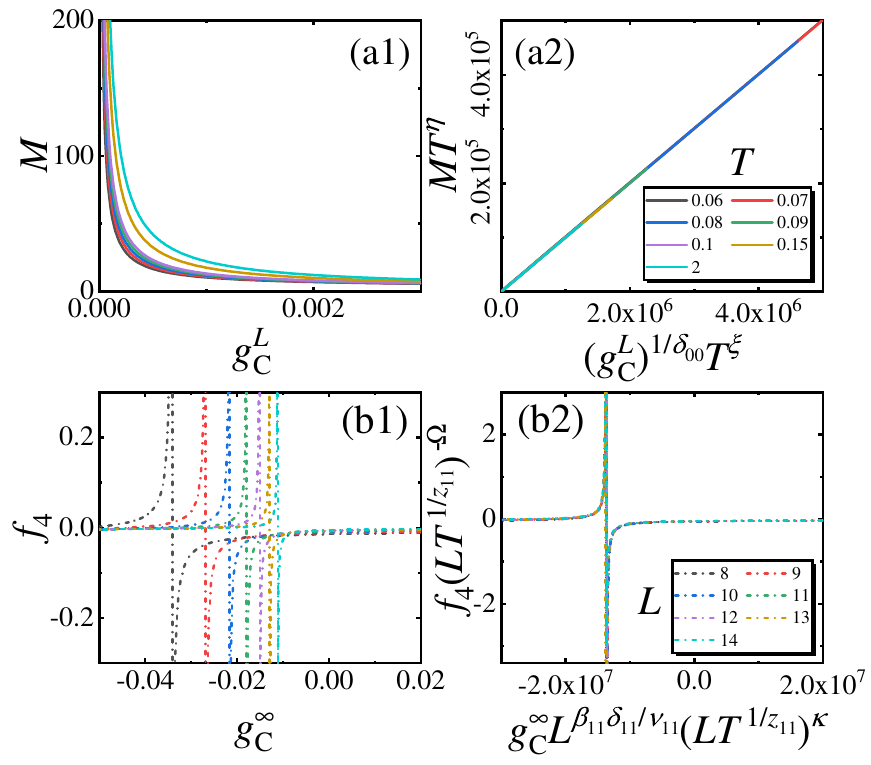}\\
  \caption{(a1) $M$ versus $g_{\rm C}^L$ for different $T$ at $L=10$.
  (a2) The rescaled curves according to Eq.~(\ref{constraint00-01}).
  (b1) $M$ versus $g_{\rm C}^{\infty}$ for different $L$.
  (b2) The rescaled curves according to Eq.~(\ref{constraint00-01}).
  In (a1) and (a2), we use $\lambda=5$, and the critical points of $(0+0)$ D YLES with $L=10$ are $h_{\rm C}^{L}=2.331231, 2.323860, 2.318697, 2.314940, 2.310248, 2.311584$ and $2.313444$ for $T=0.06, 0.07, 0.08, 0.09, 0.1, 0.15$ and $0.2$, respectively.
  In (b1) and (b2), we use $\lambda=5$, and the critical points of $(0+0)$ D YLES are $h_{\rm C}^{\infty}=2.297034$, and $LT^{1/z_{11}}=2.4$.}
\label{Fig:constrains0001}
\end{figure}

In the overlapping critical region constructed by the critical regions of $(0+0)$ D YLES and $(0+1)$ D YLES, the constraints between the scaling functions can be obtained
by substituting Eq.~(\ref{Eq:scaling01}) into Eq.~(\ref{Eq:scaling00}), which reads
\begin{eqnarray}
  M &=&T^{\xi-\eta}(g_{\rm C}^{L})^{\frac{1}{\delta_{00}}},
  \label{constraint00-01}
\end{eqnarray}
with $\xi={\beta_{01}}/({\nu_{01}z_{01}})$ and $\eta={\beta_{01}\delta_{01}}/({\nu_{01}z_{01}\delta_{00}})$.
Similarly, in the overlapping region constructed by the critical regions of $(1+0)$ D YLES and $(1+1)$ D YLES, the constraints between the scaling functions Eq.~(\ref{Eq:scaling10}) and Eq.~(\ref{Eq:scaling11}) reads
 \begin{eqnarray}
     \label{constraint10-11}
   &&f_{4}[g_{\rm Q}^{\infty}T^{-\frac {\beta_{11}\delta_{11}} {\nu_{11}z_{11}}}, g_{\rm Q}^{\infty}L^{\frac {\beta_{11}\delta_{11}} {\nu_{11}}}]\\ \nonumber
    &=& (LT^{\frac {1} {z_{11}}})^{\Omega}f_{2}[g_{\rm C}^{\infty}L^{\frac {\beta_{11}\delta_{11}} {\nu_{11}}}
    (LT^{\frac {1} {z_{11}}})^{\kappa}].
 \end{eqnarray}
with $\Omega={\beta_{11}}/{\nu_{11}}-{\beta_{01}}/{\nu_{01}}$ and $\kappa={\beta_{01}\delta_{01}}/{\nu_{01}}-{\beta_{11}\delta_{11}}/{\nu_{11}}$.
We found that both Eqs.~(\ref{constraint00-01}) and (\ref{constraint10-11}) contain the critical exponents of quantum and classical YLES, so they can be regarded as constraints between the scaling functions of quantum and classical YLES, linking quantum and classical YLES in this overlapping critical region.

In Fig.~\ref{Fig:constrains0001} (a1) and (a2), the constraints between the scaling functions of $(0+0)$ D YLES and $(0+1)$ D YLES are tested.
Here, we take $L=10$ as an example.
In Fig.~\ref{Fig:constrains0001} (a1), $M$ versus $g_{\rm C}^{L}$ for different $T$ are plotted.
The rescaled curves according to  Eq.~(\ref{constraint00-01}) match with each other as shown in Fig.~\ref{Fig:constrains0001} (a2), confirming Eq.~(\ref{constraint00-01}).
In Fig.~\ref{Fig:constrains0001} (b1) and (b2), the constraints between the scaling functions of $(1+0)$ D YLES and $(1+1)$ D YLES are tested.
Here, we fix $LT^{1/z_{11}}$ to a random constraint for different $L$.
In Fig.~\ref{Fig:constrains0001} (b1), $M$ versus $g_{\rm C}^{\infty}$ for different $L$ are plotted.
The rescaled curves according to  Eq.~(\ref{constraint10-11}) match with each other as shown in Fig.~\ref{Fig:constrains0001} (b2), confirming Eq.~(\ref{constraint10-11}).

Combining Figs.~\ref{sacle00}, \ref{scale0111T} (a1) and (a2), and \ref{Fig:constrains0001} (a1) and (a2), we have verified the applicability of the hybrid scaling mechanism in the overlapping critical region of $(0+0)$ D classical YLES and $(0+1)$ D quantum YLES.
Similarly, combining Figs.~\ref{sacle10}, \ref{scale0111T} (b1) and (b2), and \ref{Fig:constrains0001} (b1) and (b2), we have also verified the applicability of the hybrid scaling mechanism in the overlapping critical region of $(1+0)$ D classical YLES and $(1+1)$ D quantum YLES.
To the best of our knowledge, our results are the first to verify the hybrid scaling mechanism between classical and quantum phase transitions.
Moreover, since Eq.~(\ref{constraint00-01}) and Eq.~(\ref{constraint10-11}) contain information about quantum phase transitions, the introduction of these hybrid mechanisms not only expands the scope of application of hybrid scaling mechanisms but also aids in extracting information about quantum phase transitions within the framework of classical phase transitions.

\section{\label{sum}summary}

In this paper, the hybrid scaling mechanism of classical and quantum YLES is applied and numerically verified.
The model of transverse Ising chain in an imaginary longitudinal field is employed.
Besides the FPPT phase transition, this model also exhibits quantum phase transition of $(0+1)$ D YLES, $(1+1)$ D YLES and classical phase transition of $(0+0)$ D YLES and $(1+0)$ D YLES.
We numerically verify the scaling mechanisms within the critical regions of $(0+1)$ D YLES, $(1+1)$ D YLES, $(0+0)$ D YLES and $(1+0)$ D YLES.
We also studied the hybrid scaling mechanism in the overlapping critical region of two Quantum YLES and two classical YLES, respectively.
More importantly, we have applied the hybrid scaling mechanism of the overlapping critical region between the classical YLES and quantum YLES.
This study not only extends the application of hybrid scaling mechanisms to the field of classical phase transitions but also provides a way to connect classical and quantum phase transitions.
The hybrid scaling mechanism in the overlapping critical region enables us to obtain information related to quantum phase transitions through experimental measurements carried out at finite temperatures.
Given the recent experimental breakthroughs in YLES, we expect that these results will be verified in future experiments.

It is worth noting that after considering classical YLES, the model of the transverse Ising chain in an imaginary longitudinal field may not only have overlapping regions composed of two critical regions but also regions where three or four critical regions coexist.
For example, in medium-sized systems, the critical regions of (0+0) D, (0+1) D, (1+1), and (1+0) D YLES might overlap simultaneously.
The study of critical behaviors in such overlapping regions can also be approached using the hybrid scaling mechanism, albeit with more careful adjustment of various parameters.
Additionally, the HKZM proposed perviously can be regarded as a hybrid scaling mechanism for studying the non-equilibrium dynamics in the overlapping critical region of quantum phase transitions~\cite{Yin2017}.
However, for the overlapping critical region composed of classical phase transition and quantum phase transition critical regions, a unified theory to characterize the non-equilibrium dynamics therein is still lacking.
In future research, we will continue to study the non-equilibrium dynamics in the overlapping critical region composed of classical and quantum phase transition critical regions.

\section*{Acknowledgments}
This work is supported by the National Natural Science Foundation of China (Grant Nos. 12274184 and 12404105), the Qing Lan Project, and the Natural Science Foundation of the Jiangsu Higher Education Institutions of China (Grant No. 24KJB140008).

\end{CJK*}

\end{document}